\documentclass{article}    

\usepackage{graphicx}

\title{\textbf{Quantum nRules}}  
\author{Richard Mould\footnote{Department of Physics and Astronomy, State University of New York, Stony Brook,
\mbox{New York} 11794-3800; http://ms.cc.sunysb.edu/\~{}rmould}}  
\date{}    
   
\begin{document}             

\maketitle              

\begin{abstract}

Quantum mechanics traditionally places the observer `outside' of the system being studied and employs the Born interpretation.  In
this and related papers the observer is placed `inside' the system.  To accomplish this, special rules are required to engage and
interpret the Schr\"{o}dinger solutions in individual measurements.  The rules in this paper (called the nRules) do not include
the Born rule that connects probability with square modulus. 

It is required that the rules allow all conscious observers to exist continuously inside the system without empirical ambiguity --
reflecting our own unambiguous and continuous experience in the universe.  This requirement is satisfied by the nRules.  They allow
both objective and observer measurements, so state reduction can occur with or without an observer being present.

\end{abstract}

\section*{Introduction}
The method of this paper differs from that of traditional quantum mechanics in that it sees the observer 
in an ontological rather than an epistemological context.  Traditional or standard quantum theory (i.e., Copenhagen) places the
observer outside of the system where operators and/or operations are used to obtain information about the system.  This is the
epistemological model shown in \mbox{fig.\ 1}.  The observer cannot here make continuous contact with the \mbox{system -- only}
instantaneous contact.

\begin{figure}[t]
\centering
\includegraphics[scale=0.8]{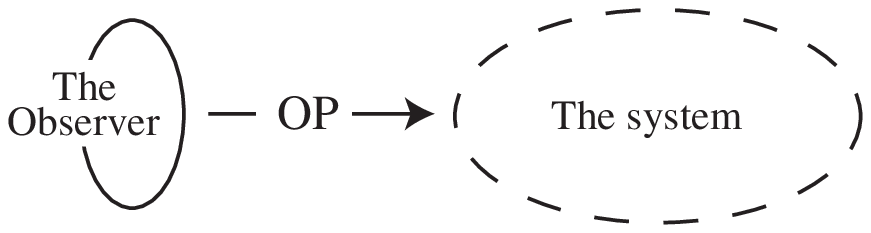}
\center{Figure 1: Epistemological Model (Copenhagen)}
\end{figure}

The large OP in fig.\ 1 might be a mathematical `operator' or a corresponding physical `operation'.  The observer makes a
measurement by choosing a formal operator that is associated with a chosen laboratory operation.  As a result, the observer is
forever outside of the observed system -- making operational choices.  The observer is forced to act apart from the system as one
who poses theoretical and experimental questions to the system.  This model is both useful and epistemologically sound.  

	However, the special rules developed in this paper apply to the system by itself, independent of the possibility that an observer
may be inside, and disregarding everything on the outside.  This is the ontological model shown in \mbox{fig.\ 2}.

\begin{figure}[h]
\centering
\includegraphics[scale=0.8]{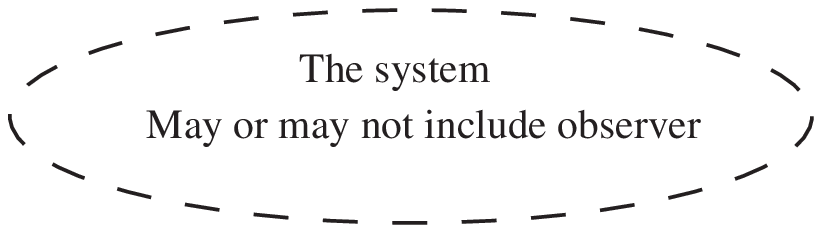}
\center{Figure 2: Ontological Model (requires special rules)}
\end{figure}

A measurement occurring inside this system is not represented by a formal operator.  Rather, it is represented by a measuring
device that is itself part of the system.  If the sub-system being measured is $S$ and a detector is $D$, then a measurement
interaction is given by the entanglement $\Phi = SD$.  If an observer joins the system in order to look at the detector, then the
system  becomes $\Phi = SDB$, where $B$ is the brain state of the observer.  Contact between the observer and the observed is
continuous in this case.

The ontological model is able to place the observer inside the universe of things and give a full account of his conscious
experience there.  It is a more realistic view of the relationship between the observer and the rest of the universe, inasmuch as
a conscious observer is always `in principle' includable in a wider system. The ontological model is a departure from the
traditional theory and has three defining characteristics:     (1) It includes observations given by $\Phi = SDB$, (2) it allows
conscious observations to be continuous, and (3) it rejects the Born rule.  In place of the Born rule, special rules like the
nRules of this paper allow physics to unambiguously predict the continuous experience of all observers in the system, including
the primary observer.  

\pagebreak

Quantum mechanical measurement is sometimes said to refer to ensembles of observations but not to individual observations.  In
this paper we propose a set of four special rules that apply to individual measurements in the ontological model.  They are called
\emph{nRules (1-4)}, and do not include the long-standing Born interpretation of quantum mechanics.  Instead, probability is
introduced (only) through the notion of \emph{probability current}.  These rules can describe a quantum mechanical state
reduction (i.e., stochastic reduction or collapse) that is associated with either an `observer' type measurement, or an
`objective' type measurement.  The former  occurs only in the presence of an observer, whereas the latter takes place
independent of an observer.  The nRules are demonstrated below in several different physical situations.  I claim that they are a
consistent and complete set of rules that can give an ontological description of any individual measurement or interaction in
quantum mechanics.  

These rules are not themselves a formal theory of measurement.  I make no attempt to understand \emph{why} they work, but strive
only to insure that they do work.   Presumably, a formal theory can one day be found to
explain these rules in the same way that atomic theory now explains the empirically discovered rules of atomic spectra, or in the way
that current theories of measurement aspire to merge with standard quantum mechanics, or to make the neurological connection with
conscious observation.

\section*{Another Rule-Set}
Other papers \cite{RM1, RM2, RM3} propose another set of rules called the \emph{oRules (1-4)}.  These are similar to the nRules
except that the basis states of reduction are confined to observer brain states -- reflecting the views of Wigner and von Neumann. 
Like the nRules, they introduce probability through the notion of `probability current' rather than through square modulus, and they
address the state reduction of conscious individuals in an ontological context -- so they give us an alternative quantum friendly
ontology.  In ref.\ 3 they are called simply the 
\mbox{\emph{rules (1-4)}}.

\section*{Purpose of Rule-Sets}
It is possible to have an empirical science using the epistemological model without explicitly talking about consciousness.  
This is because it is  assumed that the outside observer is conscious, so there is no need to make a theoretical
point of it. 
 
However, in the ontological model, everything that exists is in principle includable in the system.  So if quantum mechanics is to
be an empirical science,  the system must provide for the existence of conscious brains that can make empirical
observations.  This means that the theory must be told when and how conscious brain states appear so that an empirical science is
possible.  This cannot be done within the constrains of standard quantum mechanics. 

I emphasize again that these rule-sets are not alternative theories that seek to replace the statistical formalism of von Neumann. 
Each applies to individuals, and is like an empirical formula that requires a wider theoretical framework in order to be
understood -- a framework that is presently unknown.  I do not finally choose one of the rule-sets or propose an explanatory
theory.  I am only concerned here with the ways in which  state reduction might work in each case.

\section*{The Interaction: Particle and Detector}
Before introducing an observer into this ontological model, consider an interacting particle and detector system by itself.  These
two objects are assumed to be initially independent and given by the equation 
\begin{equation}
\Phi(t)=exp(-iHt)\psi_i\otimes d_i
\end{equation}
where $\psi_i$ is the initial particle state and $d_i$ is the initial detector state.  The particle is then allowed to pass over
the detector, where the two interact with a cross section that may or may not result in a capture.  After the interaction begins
at a time $t_0$, the state is an entanglement in which the particle variables and the detector variables are not separable.  

However, we let $\Phi(t \ge t_0)$ be in a representation whose components can be grouped so that the first term includes the
detector $d_0$ in its ground state prior to capture, and the second, third, and fourth terms describe the detector in various states
of capture  given by $d_w$, $d_{m}$, and $d_d$.  
\begin{equation}
\Phi(t \ge t_0) =\psi(t)d_0 + d_w(t) \rightarrow d_{m}(t) \rightarrow d_d(t)
\end{equation}
where $d_w(t)$ represents the entire detector immediately after a capture when only the window side of the detector is affected, and
$d_d(t)$ represents the entire detector when the result of a capture has worked its way through to the display side of the
detector.   The middle state $d_{m}(t)$ represents the entire detector during stages in between when the effects of the capture
have found their way into the interior of the detector but not as far as the display.   

\pagebreak

There is a clear discontinuity or ``quantum jump" between the two components $d_0$ and $d_w$ at the detector's window interface. 
This discontinuity is represented by a ``plus" sign and can only be bridged by a stochastic leap.  The remaining evolution from
$d_w(t)$ to $d_d(t)$ is connected by ``arrows"  and is continuous and classical.  It is  a \emph{single
component} that develops  in time and may sometimes be represented by  $d_1(t) = d_w(t) \rightarrow d_{m}(t)
\rightarrow d_d(t)$, so
\begin{equation}
\Phi(t \ge t_0)=\psi(t)d_0 + d_1(t) 
\end{equation}
The capture state $d_1(t)$ in eq.\ 3 is equal to zero at $t_0$ and increases with 
time\footnote{Each component in Eq.\ 3 has an attached environmental term $E_0$ and $E_1$ that is not shown.  These are orthogonal
to one another, insuring local decoherence.  But even though Eq.\ 3 may be decoherent locally, we will assume that the macroscopic
states $d_0$ and $d_1$ are fully coherent when $E_0$ and $E_1$ are included.  So Eq.\ 3 and others like it in this paper are
understood to be coherent when universally considered.  We will call them ``superpositions", reflecting their global rather than
their local properties.}$^,$\footnote{Superpositions of macroscopic states have been found at low temperatures \cite{JRF}.  The
components of these states are locally coherent for a measurable period of time.  }. The state $\psi(t)$ is a free particle as a
function of time, including all the incoming and scattered components.  It does no harm and it is convenient to let $\psi(t)$ carry
the total time dependence of the first component, and to let $d_0$ be normalized throughout\footnote{Equation  3 can be written with
coefficients
$c_0(t)$ and $c_1(t)$ giving $\Phi(t \ge t_0) = c_0(t)\psi(t)d_0 + c_1(t)d_1$, where the states $\psi(t)$, $d_0$, and $d_1$ are
normalized throughout.  We let
$c_0(t)\psi(t)$ in this expression be equal to $\psi(t)$ in eq.\ 3, and let $c_1(t)d_1$ be equal to $d_1(t)$ in eq.\ 3.}.

The first component in eq.\ 3 is a superposition of all possible scattered waves of $\psi(t)$ in product with all possible recoil
states of the ground state detector, so $d_0$ is a spread of detector states including all the recoil possibilities together with
their correlated environments.  The second component is also a  superposition of this kind.  It includes all of the recoil
components of the detector that have captured the particle.

\section*{Add an Observer}
Assume that an observer is looking at the detector in eq. 1 from the beginning.  
\begin{displaymath}
\Phi(t)=exp(-iHt)\psi_i\otimes D_iB_i
\end{displaymath}
where $B_i$ is the observer's initial brain state that is entangled with the detector $D_i$.  This brain is understood to
include \emph{only} higher order brain parts -- that is, the physiology of the brain that is directly associated with consciousness
after all image processing is complete.  All lower order physiology leading to $B_i$ is assumed to be part of the detector.  The
detector is now represented by a capital $D$, indicating that it includes the bare detector by itself \emph{plus} the low-level
physiology of the observer.

Following the interaction between the particle and the detector, we  have
\begin{eqnarray}
\Phi(t \ge t_0) &=& \psi(t)D_0B_0 + D_w(t)B_0 \rightarrow D_{m}(t)B_0 \rightarrow D_d(t)B_1 \hspace{.5cm}  \\
\mbox{or} \hspace{.5cm} \Phi(t \ge t_0) &=& \psi(t)D_0B_0 + D_1(t)B_1 \nonumber
\end{eqnarray}
where $B_0$ is the observer's brain when the detector is observed to be in its ground state $D_0$, and $B_1$ is the brain when the
detector is observed (at the display end) to be in its capture state $D_1$.  Since the
detector now includes the low-level physiology of the observer's brain, the display end of the detector $D_d$ is the interface between
 lower level and upper level physiology.  As before, a discontinuous quantum jump is represented by a plus sign, and the continuous
 evolution of a single component is represented by an arrow.  

If the interaction is long lived compared to the time it takes for the signal to travel through the
detector in eq.\ 4, then the superposition in that equation might  exist for a long time before a capture causes a
state reduction.  This means that there can be two active brain states of this observer in superposition, where one sees the
detector in its ground state and the other simultaneously sees the detector in its capture state.   Equation 4 therefore invites a
paradoxical interpretation like that associated with Schr\"{o}dinger's cat.  This ambiguity cannot be allowed.  The nRules of this
paper must not only provide for a stochastic trigger that gives rise to a state reduction, and describe that reduction, they must also
insure than an empirical ambiguity of this kind will not occur.

\section*{The nRules}
	The first rule establishes the existence of a stochastic trigger.  This is a property of the system that has nothing to do with
the kind of interaction taking place or its representation.  Apart from making a choice, the trigger by itself has no effect on
anything.  It initiates a state reduction only when it is combined with nRules 2 and 3.

\vspace{.4cm}

\noindent
\textbf{nRule (1)}: \emph{For any subsystem of n complete components in a system having a total square modulus equal to s, the
probability per unit time of a stochastic choice of one of those components at time t is given by $(\Sigma_nJ_n)/s$, where
the net probability current $J_n$ going into the $n^{th}$ component at that time is positive.}

\noindent
[\textbf{note}: A \emph{complete component} is a solution of Schr\"{o}dinger's equation that includes all of the (symmetrized)
objects in the universe.  It is made up of \emph{complete states} of those objects  including all their state variables.  If
$\psi(x_1, x_2)$ is a two particle system with inseparable variables $x_1$ and $x_2$, then $\psi$ is considered to be a single 
object.  All such  objects are included in a complete component.  A component that is a sum of
less than  the full range of a variable (such as a partial Fourier expansion) is not complete.]

\vspace{.4cm}

The second rule specifies the conditions under which \emph{ready states} appear in solutions of Schr\"{o}dinger's equation.
These are understood to be the basis states of a state 
reduction.

\vspace{.4cm}

\noindent
\textbf{nRule (2)}: \emph{If an irreversible interaction produces complete components that are discontinuous with the initial 
component, then all of the new states that appear in these components will be ready states.}

\noindent
[\textbf{note}: Continuous means continuous in all variables. Although solutions to Schr\"{o}dinger's equation change
continuously in time, they can be \emph{discontinuous} in other variables -- e.g., the separation between the $n^{th}$ and the $(n +
1)^{th}$ orbit of an atom with no orbits in between.  A discontinuity  can also
exist between macroscopic states that are locally decoherent.   For instance, the displaced detector states $d_0$ (ground state) and
$d_1$ (capture state) appearing in \mbox{eq.\ 3} are discontinuous with respect to detector variables.  There is no state $d_{1/2}$
in between.  Like atomic orbits, these detector states are a `quantum jump' apart.]

\noindent
[\textbf{note}: The \emph{initial component} is the first complete component that appears in a given solution of Schr\"{o}dinger's
equation.  A  solution is defined by a specific set of boundary conditions.  So eqs.\ 1 and 3 are both included in the single
solution that contains the discontinuity between $d_0$ and $d_1$, where eq.\ 1 (together with its complete environment) is the
initial state.  However, boundary conditions change with the collapse of the wave function.  The single  component that
survives a collapse will be complete, and will be the initial component of the new solution.]

\vspace{.4cm}

The collapse of a wave function and the change of a ready state to a \emph{realized} state is provided for by nRule (3).  If a
complete state is not `ready' it will be called `realized'.  We therefore introduce dual state categories where ready
states are the basis states of a collapse.  They are on stand-by, ready to be stochastically chosen and converted by nRule (3) to
realized states.  In this paper, ready states are underlined and realized states are not.    

\vspace{.4cm}

\pagebreak

\noindent
\textbf{nRule (3)}: \emph{If a component is stochastically chosen during an interaction, then all of the ready states that result
from that interaction (using nRule 2) and that are unique to that component will become realized, and all other components in
tthe superposition will be immediately reduced to zero.}

\noindent
[\textbf{note}: The claim of an immediate (i.e., discontinuous) reduction is the simplest possible way to describe the collapse
of the state function.  A collapse is brought about by an instantaneous change in the boundary conditions of the
Schr\"{o}dinger equation, rather than by the introduction of a new `continuous' mechanism of some kind.  A continuous
modification can be added later (with a  modification of nRule 3) if that is seen to be 
necessary\footnote{The new boundary comes from a stochastic hit on one of the available eigenvalues, which \emph{is} the new
boundary.  The stochastic trigger is intrinsically discontinuous and imposes that discontinuity on the developing wave
function.  }.]

\noindent
[\textbf{note}:  This collapse does not generally preserve normalization.  That does not alter probability in subsequent
reductions because of the way probability per unit time is defined in nuRule (1) -- that is, divided by the total square modulus.]

\vspace{.4cm}

Only positive current going into a \emph{ready component} (i.e., a component containing ready states) is physically meaningful
because it represents positive probability.  A negative current (coming out of a ready component) is not physically meaningful and
is not allowed by nRule (4).  Without this restriction, probability current might flow in-and-out of one ready component and into
another.  The same probability current would then be `used' and `reused', and this would generally distort the total probability
of a process. To prevent this we say     

\vspace{.4cm}

\noindent
\textbf{nRule (4)}: \emph{A ready component cannot transmit probability current.}

\vspace{.4cm}

If an interaction does not produce  complete  components that are discontinuous with the initial component, then the
Hamiltonian will develop the state in the usual way, independent of these rules.  If the stochastic trigger selects a component
that does not contain ready states, then there will be no nuRule (3) state reduction.

\pagebreak

\section*{An Example}
A free neutron decay is written $\Phi = n + \underline{ep\overline{\nu}}$, where the second component is zero at $t = 0$ and
increases in time as probability current flows into it.  This component contains three entangled particles  making a whole  object,
where all three  are `ready' states as indicated by the underline (see nRule 2). Each component of $\Phi$ is multiplied by a term
representing the environment (not shown).  Each  is complete for this reason and also because the variables of each particle
take on all of the values that are allowed by the boundary conditions.  Following \mbox{nRule (3)} there will be a
stochastic hit on $\underline{ep\overline{\nu}}$, reducing the system to the realized correlated states $ep\overline{\nu}$.  

Specific values of, say, the electron's momentum are not stochastically chosen by this reduction, because all possible values are
included in $ep\overline{\nu}$.  For detail of this kind, a detector must be added that measures the magnitude of the electron's
momentum in a specific direction away from the chosen decay site.  That will require another stochastic hit on the component that
includes that detector.   A detector interaction and reduction of this kind is discussed in the next section.

\section*{Apply Rules to Detector Interaction}
To see how the nRules carry out a reduction that involves an observed detector, we apply them to the first row of eq.\ 4.  This only
affects the first two components
\begin{equation}
\Phi(t \ge t_0)=\psi(t)D_0B_0 + \underline{D}_w(t)B_0 
\end{equation}
where the second component $\underline{D}_w(t)B_0$ contains a ready state (underlined) by virtue of nRule (2).  Components
$\underline{D}_m$ and
$\underline{D}_d$, do not appear this equation because nRule (4) will not allow $\underline{D}_w$ to pass current to them.  Since the
second brain state $B_0$ in eq.\ 5 is the same as the first, there is only one brain state in the superposition.  A cat-like ambiguity
is thereby avoided.    Equation 5 now
\emph{replaces} eq.\ 4.   

	Equation 5 is the state of the system before there is a stochastic hit that produces a state reduction.  The observer is here
consciously aware of the detector in its ground state $D_0$.  If there is a capture, then there will be a stochastic hit on the
second component of eq.\ 5 at a time $t_{sc}$. This will reduce the first component to zero according to nRule (3), and convert
the ready state in the second component to a realized state.
\begin{displaymath}
\Phi(t = t_{sc} > t_0) = D_w(t)B_0
\end{displaymath}
The observer is still conscious of the detector's ground state in this equation because the capture has only affected the window end
of the detector.  But after $t_{sc}$, a continuous evolution will produce
\begin{equation}
\Phi(t \ge t_{sc} > t_0) = D_w(t)B_0 \rightarrow D_{m}(t)B_0 \rightarrow D_d(t)B_1
\end{equation}
Since this equation represents a single component that evolves in time as shown, there is no time at which both $B_0$ and $B_1$
appear simultaneously.  There is therefore no cat-like ambiguity in eq. 6.

	Standard quantum mechanics (without these rules) gives us eq.\ 4 by the same logic that it gives us Schr\"{o}dinger's cat and
Everett's many worlds.  Equation 4 (top or bottom row) is a single equation that simultaneously presents two different conscious
brain states, resulting in an unacceptable ambiguity.  However with these nRules in effect, the Schr\"{o}dinger solution is
properly grounded in observation, allowing the rules to correctly and unambiguously predict the experience of the observer.  This
is accomplished by replacing `one' equation in eq.\ 4 with `two' equations in eqs.\ 5 and 6.   Equation 5 describes the state of
the system \emph{before} capture, and eq.\ 6 describes the state of the system \emph{after} capture.  Before and after are two
different solutions to Schr\"{o}dinger's equation, specified by different boundary conditions.  Remember, we said that the
stochastic trigger selects the (new) boundary that applies to the reduced state.  So it is the stochastic event that
separates the two solutions Ð- defining before and after.  

	If there is no stochastic hit on the second component in eq.\ 5, then it will become a \emph{phantom} component.  A component is
a phantom when there is no longer any probability current flowing into it (in this case because the interaction is complete), and
when there can be no current flowing out of it because it includes ready states that comply with nRule (4).  The phantom
properties of a component extends to the complete component.   A phantom component can be dropped out of the equation without
consequence.  Doing so only changes the definition of the system.  It is the same kind of redefinition that occurs in standard
practice when one chooses to renormalize a system at some new starting time.  Keeping a phantom is like keeping the initial system. 
Because of nRule (3), kept phantoms are reduced to zero whenever another component is stochastically chosen.
  
With no stochastic hit in eq.\ 5,  the new system (dropping the phantom $\underline{D}_wB_0$) is just the first component
of that equation.  This corresponds to the observer continuing to see the ground state detector $D_0$, as he should in this case.

\section*{A Terminal Observation}
	An observer who is inside a system must be able to confirm the validity of the Born rule that is normally applied from the
outside.  To show this, suppose our observer is not aware of the detector during the interaction with the particle, but
he looks at the detector after  a time $t_f$ when the primary interaction is
complete.  During the interaction we have 
\begin{equation}
\Phi(t_f > t \ge t_0) = [\psi(t)d_0 + \underline{d}_w(t)]\otimes X
\end{equation}
where $X$ is the unknown state of the observer prior to the physiological 
interaction\footnote{The ``decision" of the observer to look at the detector is assumed to be deterministically internal in the
ontological model.  In this respect, the ontological model is like classical physics.}. 

Assume there has \emph{not} been a capture.  Then after the interaction is complete and before the observer looks at the detector
we have
\begin{displaymath}
\Phi(t \ge t_f > t_0) = [\psi(t)d_0 + \underline{d}_w(t_f)]\otimes X
\end{displaymath}
where there is no longer a probability current flow inside the bracket. The second component in the bracket is therefore a
phantom.  There is no current flowing into it, and none can flow out of it because of nRule (4).  So the equation is
essentially\footnote{Again, deciding to drop $\underline{d}_w(t_f)$ amounts to redefining the system from a new starting time.}. 
\begin{displaymath}
\Phi(t \ge t_f > t_0) = \psi(t)d_0\otimes X
\end{displaymath}
When the observer finally observes the detector at $t_{ob}$ he will get
\begin{displaymath}
\Phi(t \ge t_{ob} > t_f > t_0) = \psi(t)d_0\otimes X \rightarrow \psi(t)D_0B_0 
\end{displaymath}
where the physiological process (represented by the arrow) carries $\otimes X$ into $B_0$ and $d_0$ into $D_0$ by a continuous
classical progression leading from independence to entanglement.   This corresponds to the observer coming on board to witness
the detector in its ground state as he should when is no capture.  The probability of this happening is equal to the square
modulus of $\psi(t)d_0\otimes X$ in eq.\ 7.

If the particle \emph{is} captured during the primary interaction, there will be a stochastic hit on the second component inside
the bracket of eq.\ 7 at some time $t_{sc} < t_f$.  This results in a capture given by
\begin{displaymath}
\Phi(t_f > t = t_{sc} > t_0) = d_w(t)\otimes X
\end{displaymath}
after which 
\begin{displaymath}
\Phi(t_f > t \ge  t_{sc} > t_0) = [d_w(t) \rightarrow d_m(t) \rightarrow d_d(t)]\otimes X = d_1(t)\otimes X
\end{displaymath}
as a result of the classical progression inside the detector.  When the observer does become aware of the detector at $t_{ob} > t_f$
we finally get
\begin{displaymath}
\Phi(t \ge t_{ob} > t_f > t_{sc} > t_0) = d_1\otimes X \rightarrow D_1B_1
\end{displaymath}
So the observer comes on board to witness the detector in its capture state with a probability equal to the square modulus of
$\underline{d}_w(t_f)\otimes X$ in eq.\ 7.  The nRules therefore confirm the Born rule, in this case as a theorem.

\section*{An Intermediate Case}
	In eq.\ 5 the observer is assumed to interact with the detector from the beginning.  Suppose that the incoming particle results
from a long half-life decay, and that the observer's physiological involvement only \emph{begins} in the middle of the primary
interaction.  Before that time we will have
\begin{displaymath}
\Phi(t \ge  t_0) = [\psi(t)d_0 + \underline{d}_w(t)]\otimes X
\end{displaymath}
where again $X$ is the unknown brain state of the observer prior to the physiological interaction.  Primary probability current
here flows between the detector components inside the bracket.  

Let the observer look at the detector at some time $t_{look}$ giving 
\begin{eqnarray}
\Phi &=& \psi(t_{look})d_0\otimes X \rightarrow \psi(t_{look} + \pi)D_0B_0  \\
&+& \underline{d}_w(t_{look})]\otimes X \rightarrow \underline{D}_w(t_{look} + \pi)B_0  \nonumber
\end{eqnarray}
where $\pi$ is the time for the observer to come on board.  The  first row of this equation is a
single component that evolves continuously and classically as represented by the arrow.  That evolution carries $\otimes X$ into
$B_0$ and $d_0$ into $D_0$ by a process that leads from independence to entanglement. 
The primary interaction is still active during this time, and this gives rise to a vertical current going from the first to the
second row in eq.\ 8.  The second row is therefore a continuum of components that are created parallel to the
first row at each moment of time.  So at  time $t_{look} + \pi$, vertical current  flows only into the final component
$\underline{D}_w(t_{look} + \pi)B_0$ in the second row of eq.\ 8. Components prior to the last one no longer have current flowing
into them from above, and since there is no horizontal current among these ready states, they become phantom components as soon as
they are created.  Therefore, when the observation is complete at the time $t_{ob} = 
t_{look} + \pi$, we can write eq. 8 as  
\begin{equation}
\Phi(t \ge t_{ob}  > t_0) = \psi(t)D_0B_0 + \underline{D}_w(t)B_0
\end{equation}
where $\underline{D}_w(t)B_0$ is zero at $t_{ob} $ and increases in time.  The system has been redefined to eliminate all
those phantoms.  This equation is identical with eq.\ 5; so from this moment on, it is as though the observer has been on board from
the beginning.  

If there is a subsequent capture at a time $t_{sc}$, this will become like eq. 6.
\begin{equation}
\Phi(t \ge t_{sc} > t_{ob} > t_0) = D_w(t)B_0 \rightarrow D_m(t)B_0 \rightarrow D_d(t)B_1
\end{equation}
  
	If there is a stochastic hit between $t_{look}$ and $t_{ob}$, then the corresponding ready state in the second row of eq.\ 8 will
be chosen and made a realized state.  It will then proceed classically and continuously to $D_dB_1$ as in eq.\ 10.

\section*{A Second Observer}
If a second observer is standing by while the first observer interacts with the detector during the primary interaction, the state
of the system will be
\begin{displaymath}
\Phi(t \ge  t_0) = [\psi(t)D_0B_0 + \underline{D}_w(t)B_0]\otimes X
\end{displaymath}
where $X$ is an unknown state of the second observer prior to his interacting with the system.  The detector $D$ here includes the
low-level physiology of the first observer.  A further expansion of the detector will include the second observer's low-level
physiology when he comes on board.  When a product of brain states appears in the form $BB$ or $B\otimes X$, the first term will
refer to the first observer and the second to the second observer.  
	
The result of the second observer looking at the detector will be the same as that found for the first observer in the previous
section, except now the first observer will be present in each case.  In particular, the equations similar to \mbox{eqs.\ 8, 9, and
10} are now 
\begin{eqnarray}
\Phi &=& \psi(t_{look})D_0B_0\otimes X \rightarrow \psi(t_{ob} )D_0B_0B_0 
\nonumber\\ &+& \underline{D}_w(t_{look})B_0\otimes X \rightarrow \underline{D}_w(t_{ob})B_0B_0  \nonumber
\end{eqnarray}
\begin{displaymath}
\Phi(t \ge t_{ob}  > t_0) = \psi(t)D_0B_0B_0 + \underline{D}_w(t)B_0B_0  
\end{displaymath}
\begin{displaymath}
\Phi(t \ge t_{sc} > t_{ob}  > t_0) = D_w(t)B_0B_0 \rightarrow D_m(t)B_0B_0 \rightarrow D_d(t)B_1B_1 
\end{displaymath}
These will all yield the same result for the new observer as they did for the old observer.  In no case will the nRules produce a
result like $B_1B_2$ or $B_2B_1$.

\vspace{.4cm}

In previous sections we have seen how the nRules go about including observers inside a system in an ontological model.  These rules
describe when and how the observer becomes conscious of measuring instruments, and replicate common empirical experience in these
situations. The nRules are also successfully applied in another paper \cite{RM5}  where two versions of the Schr\"{o}dinger cat
experiment are examined.  In the first version a conscious cat is made unconscious by a stochastically initiated process; and in the
second version an unconscious cat is made conscious by a stochastically initiated process. 

In the following sections we turn attention to another problem -- the requirement that macroscopic states must appear in
their normal sequence.   This sequencing chore represents a major application of nRule (4) that is best illustrated in the case of a
macroscopic counter.

\section*{A Counter}

If a beta counter that is exposed to a radioactive source is turned on at time $t_0$, its state function will be given by
\begin{displaymath}
\Phi(t \ge   t_0) = C_0(t) + \underline{C}_1(t)\
\end{displaymath}
where $C_0$ is a counter that reads zero counts, $C_1$ reads one count, and $C_2$ (not shown) reads two counts, etc.  The second
component $\underline{C}_1(t)$ is zero at $t_0$ and increases in time.  The underline indicates that it is a ready state as
required by nRule (2).  $C_2$ and higher states do not appear in this equation because \mbox{nRule (4)} forbids current to leave
$\underline{C}_1(t)$.  Ignore the time required for the capture effects to go from the window to the display end of the counter.  

\begin{figure}[t]
\centering
\includegraphics[scale=0.7]{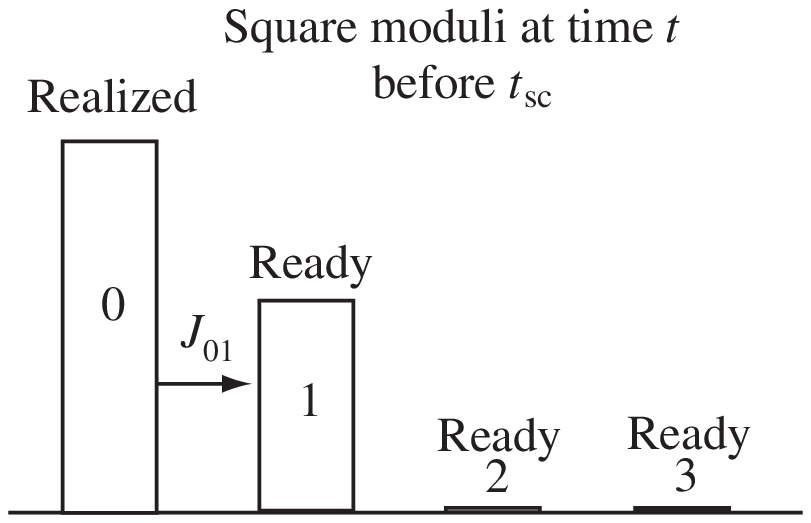}
\center{Figure 3}
\end{figure}

Therefore, the $0^{th}$ and the $1^{st}$ components are the only ones that are initially involved, where the current flow is $J_{01}$
from the $0^{th}$ to the $1^{st}$ component.  The resulting distribution at some time $t$ before $t_{sc}$ is shown in fig.\ 3,
where
$t_{sc}$ is the time of a stochastic hit on the second component.

This means that the $1^{st}$ component \emph{will} be chosen because all of the current from the (say,
normalized) $0^{th}$ component will pore into the $1^{st}$ component making $\int J_{01}dt = 1.0$.  Following the stochastic hit
on the $1^{st}$ component, there will be a collapse to that component because of \mbox{nRule (3)}.  The first two dial readings
will therefore be sequential, going from 0 to 1 without skipping a step such as going directly from 0 to 2.  It is nRule (4) that
enforces the no-skip behavior of macroscopic objects, for without it any component in the superposition might be chosen.    It is
empirically mandated that macroscopic states should always follow in sequence without skipping a step.

With the stochastic choice of the $1^{st}$ component at $t_{sc}$, the process will begin again as shown in 
fig.\ 4.  This also leads with certainty to a stochastic choice of the $2^{nd}$ component.  That certainty is accomplished by the
wording of nRule (1) which requires that the probability per unit time is given by the current flow $J_{12}$ divided by the total
square modulus at that moment.  The total integral $\int J_{12}dt$  is less than 1.0 in  fig.\ 4 because of the reduction that
occurred in fig.\ 3;  but it is restored to 1.0 when divided by the total square modulus.  It is therefore certain that the
$2^{nd}$ component will be chosen.

\begin{figure}[h]
\centering
\includegraphics[scale=1.2]{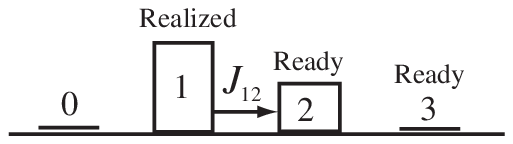}
\center{Figure 4}
\end{figure}

And finally, with the choice of the $2^{nd}$ component, the process will resume again with current $J_{23}$ going from the
$2^{nd}$ to the $3^{rd}$ component.  This leads with certainty to a stochastic choice of the $3^{rd}$ component.  

 If an observer watches the counter from the beginning he will be able to see it go sequentially from $C_0$ to $C_1$, to $C_2$,
etc., for he is himself a continuous part of the system.  He does not `peek' intermittently like an epistemological observer who is
\emph{not} part of the system -- relating to it only through the Born rule.  Although the empirical experience of the ontological
observer is different from the epistemological observer, there is no contradiction between the two possibilities.  The ontological
observer gets `continuous' information about the behavior of the counter, whereas the epistemological observer sees only
discontinuous parts of the (same) picture.

\section*{The Parallel Case} 

Now imagine parallel states in which a quantum process may go either clockwise or counterclockwise as shown in Fig. 5.  Each
component includes a macroscopic piece of laboratory apparatus $A$, where the Hamiltonian provides for a discontinuous clockwise
interaction going from the $0^{th}$ to the $r^{th}$ state, and another one going from there to the final state $f$; as well as a
comparable counterclockwise interaction from the $0^{th}$ to the $l^{th}$ state and from there to the final state $f$.  The
Hamiltonian does not provide a direct route from the $0^{th}$ to the final state, so the system will choose stochastically between a
clockwise and a counterclockwise route. 

\begin{figure}[h]
\centering
\includegraphics[scale=0.8]{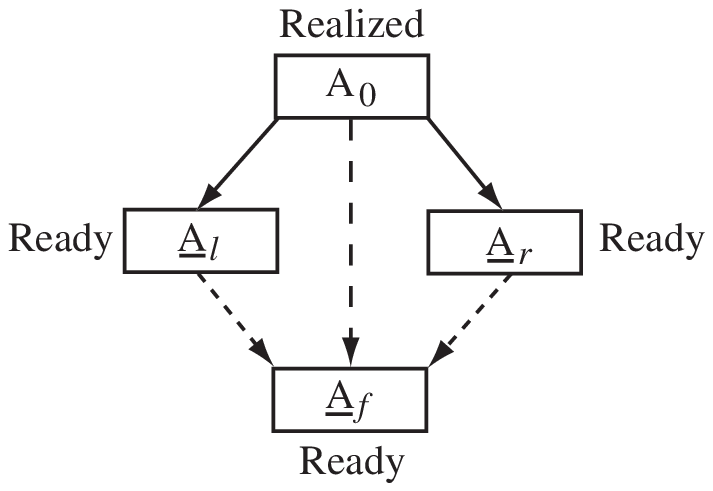}
\center{Figure 5}
\end{figure}

With nRule (4) in place, probability current cannot initially flow from either of the intermediate states to the final state, for
that would carry a ready state into another  state.  The dashed lines in fig.\ 5 indicate the forbidden
transitions.  But once the state $\underline{A}_l$ (or $\underline{A}_r$) has been stochastically chosen, it will become a
realized state $A_l$ (or $A_r$) and a subsequent transition to $\underline{A}_f$ can occur that realizes $A_f$.  

The effect of nRule (4) is therefore to force this macroscopic system into a classical sequence that goes either clockwise or
counterclockwise.  Without it, the system might make a  second order transition (through one of the intermediate states) to the final
state, without the intermediate state being realized.  The observer would then see the initial state followed by the final state,
without knowing (in principle) which pathway was followed.  This is familiar behavior of standard quantum mechanical microscopic
systems, but it should not be found in macroscopic systems. In Heisenberg's famous example formalized by Feynman, a microscopic
particle observed at point $a$ and later at point $b$ will travel over a quantum mechanical superposition of all possible paths in
between.  Without nRule (4), macroscopic objects facing discontinuous parallel choices would do the same thing.  The fourth nRule
therefore forces the parallel system into one or the other classical path, so it is not a quantum mechanical superposition of both
paths.

\section*{A Continuous Variable}
In the above examples nRule (4) guarantees that sequential discontinuous steps in a superposition are not passed over. 
If the variable itself is classical and continuous, then continuous observation is possible without the
necessity of stochastic jumps.  In that case we do not need \mbox{nRule (4)} or any of the \mbox{nRules (1-4)}, for they do not
prevent or in any way qualify the motion.  

However, a classical variable may require a quantum mechanical jump-start.  For instance, the
mechanical device that is used to  seal the fate of Schr\"{o}dinger's cat (e.g., a falling hammer) begins its motion with a
stochastic hit.  That is, the decision to begin the motion (or not) is left to a $\beta$-decay.  In this case \mbox{nRule (4)}
forces the motion to begin at the beginning, insuring that no value of the classical variable is passed over;  so the hammer
will fall from its \emph{initial} angle with the horizontal.   Without nRule (4), the hammer might begin its fall at some other angle
because probability current will flow into angles other than the initial one.  With nRule (4) in place, no angle will be passed
over.

\section*{Microscopic Systems}
The discussion so far has been limited to experiments or procedures whose outcome is commonly known.  Our claim has been that the
nRules are chosen to work without regard to a theory as to `why' they work.  Therefore, with the exception of our example of a free
neutron decay, our attention has always gone to macroscopic situations in which the results are directly available to conscious
experience.  However, if the nRules are correct we should also want know how they apply to microscopic systems, even though the
predicted results in these cases are more speculative.  In this section we will consider the implications of the nRules in two
microscopic cases.  The important question to ask in each case is: Under what circumstances will these rules result in a state
reduction of a microscopic system?

\pagebreak

\begin{displaymath}
\mbox{\textbf{Case 1 -- spin states}}
\end{displaymath}

No state reduction will result from changing the representation.  In particular, replacing the spin state $+z$ with the sum of states
$+x$ and $-x$ will not result in either one of the $x$-states becoming a ready state.  

	This will be true for a spin $+z$ particle even if the common environment $E$ includes a magnetic field that is continuously
applied in (say) the $x$-direction.  So long as the magnetic field  is the same for both $+x$ and $-x$, the result will
be the same -- i.e., neither one will become a ready state.  

If any part of the environment interacts differently with $+x$ than it does \mbox{with $-x$}, decoherence will occur
over time.  In that case we will have continuously changing environments $E^p(t)$ and $E^n(t)$, such that
\begin{equation}
\Phi(\tau > t \ge t_0) = \frac{1}{\sqrt{2}}(+z)E^p(t) + \frac{1}{\sqrt{2}}(-z)E^n(t)
\end{equation}
where $E^p(0) = E^n(0) = E$, and $E^p(\tau) = E_+,  E^n(\tau) = E_-$.  States $E_+$ and $E_-$ are the final
environments of $(+x)$ and $(-x)$ respectively, and the time $\tau$ will be either the time at which  decoherence  is
complete, or the time at which the components have become decoherent to some other desired extent.  It may be that the environments
$E_+$ and $E_-$ are \emph{similar} to $E$ and to each other (i.e., same temperature and pressure, same particles, same radiation
field, etc.), but time will change $E_+$ and $E_-$ so they can no longer be \emph{identical} with $E$ or with each other.  Of
course, it might also be that $E_+$ and $E_-$ are not  even similar to $E$.

Taking the first and last term in eq.\ 11 we write
\begin{eqnarray}
\Phi(\tau > t \ge t_0) &=& \frac{1}{2}[(+z) + (-z)]E \rightarrow \frac{1}{2}[(+z)E_+ + (-z)E_-] \\
&&   \hspace{1cm}  t =t_ 0  \hspace{2.8cm} t = \tau  \nonumber
\end{eqnarray}
where the arrow indicates a continuous transition.  The second square bracket now differs from the first in that it is partially or
fully decoherent.  But since $E$ goes continuously into $E_+$ or $E_-$, nRule (2) says that no `ready' states will result. 
Therefore, there can be \emph{no state reduction} in this case.  According to the nRules, decoherence does not lead to a collapse
of the wave to either $+x$ \mbox{or $-x$} because decoherence is not discontinuous.  

	If the magnetic field is non-homogeneous there will be a physical separation of the $+x$ and $-x$ states, but that will not change
the above analysis or its conclusion. Assuming that this field is continuously (i.e., classically) applied,  there  will   be no
state reduction.  The separation of the + and - states only means that $E_+$ and $E_-$ are more likely to be different, and that
decoherence will probably come closer to completion.  Ready states will appear only when the $+x$ and $-x$ components are picked up
by different detectors at different locations, resulting in a \emph{detector related} discontinuity.  According to nRule (3), a
state reduction can only occur if and when that happens.

	There are many other `quantum' processes that are continuous and therefore not subject to stochastic reduction, such
as scattering, interference, diffraction, and tunneling.  

\begin{displaymath}
\mbox{\textbf{Case 2 -- atomic emission}}
\end{displaymath}

	Imagine that a single atom decays from its first excited state $A_1$ to its ground state plus a photon $\gamma A_0$.  There
is a discontinuity between the two states, for the system does not go continuously through atomic orbits between $A_1$ and $A_0$. 
The photon is either released or it isn't.  Therefore,  
\begin{displaymath}
\Phi(t \ge t_0) = A_1(t) + \underline{\gamma A}_0(t)
\end{displaymath}
where the decay is said to begin at $t_0$.  The second component is zero at $t_0$ and increases in time.  As the first component
$A_1$ goes to zero all of the current will flow into the second component, insuring that there will be a stochastic hit on
$\underline{\gamma A}_0(t)$ at some time $t_{sc}$.  When that happens we will have
\begin{equation}
\Phi(t \ge t_{sc} > t_0) = \gamma (t)A_0(t_{sc})
\end{equation}
where $\gamma (t)$ is a photon pulse that leaves the atom at time $t_{sc}$. The role of \mbox{nRule (4)} in this equation is to block
the release of the photon until the atom has been stochastically chosen.  That is not the way of standard quantum mechanics, where the
photon field includes radiation from all possible moments of release prior to $t_{sc}$.  

This means that the moments of decay and photon emission are stochastically chosen.  If an atom somewhere in the universe releases a
photon, the time of its release is made definite by the nRules, whether or not the photon is ever detected or observed.  The same
would be true of any kind of microscopic decay.  We do not normally think of the collapse of a quantum mechanical wave as taking place
at the atomic level, but there is certainly no evidence against it.

The nRules also guarantee that the  collapse in eq.\ 13 will go no further; that is, it cannot choose the photon's \emph{direction}
of emission.  Such a choice would involve a stochastic selection from a continuum of eigenstates representing specific angular
directions.  One of these eigenstates cannot alone be a complete component because it doesn't range over all angles, so it cannot
be a `ready' component according to nRule (2).   It cannot be a basis state of reduction.  The choice of direction requires a
separate collapse that includes the identification of the affected detector at a specific location away from the emitting atom.  The
problem is then similar to the one previously considered except that the emitting atom must be in the picture.  That's because the
atomic recoil must be correlated (non-locally) with the detector of choice in order to insure the conservation of momentum.  

	Our conclusion following the nRules is that \emph{any} discontinuous and irreversible process (microscopic or macroscopic) is
subject to a stochastic hit and consequent state reduction.  The widely held belief that a state reduction does not occur at the
atomic level goes back to the Copenhagen belief that a collapse must be macroscopic, requiring the presence of classically sized
instruments.  Copenhagen supports the notion that there is a fundamental difference between the classical world and the quantum
world, but that is not the
  view of this paper.  The nRules endorse the idea that there are continuous and discontinuous process
at all levels -- the macroscopic and the microscopic.  Continuous processes are guided by the Schr\"{o}dinger equation and make no
use of the nRules.  Discontinuous `quantum jumps' at all levels are bridged by stochastic hits that are governed by the nRules.

\section*{Decoherence}
Suppose that two states $A$ and $B$ that are initially in coherent Rabi oscillation and are exposed to a phase disrupting
environment.  This may be expressed by adopting eq.\ 12 to give
\begin{eqnarray}
\Phi(\tau > t \ge t_0) &=& (A + B)E \rightarrow AE_A + BE_B \\
&&   \hspace{.3cm}  t = t_0  \hspace{1.6cm} t = \tau  \nonumber
\end{eqnarray}
The transition that carries $A$ into $B$ and $B$ into $A$ is reversible.  Therefore, neither one is a ready state according to nRule
(2).  The partial decoherent state $(AE_A + BE_B)$ evolves continuously in this equation from the initial state \mbox{$(A + B)E$},
so no ready states are created, and there are no possible state reductions.  Because the phase relationship between $A$ and $B$ is
continuously destroyed due to decoherence, the oscillations will diminish during this decay, approaching zero as $t$ goes to $\tau$. 

Evidence for diminishing oscillations of this kind is found in low temperature experiments \cite{DV, YY}  where oscillations
undergo decoherent decay without any sign of interruption due to stochastic state reduction. Because probability current flows
continuously back and forth between the two components, one might expect to see a state reduction within a single cycle.  But that
doesn't happen for the reasons stated above.  

\vspace{.4cm}

Equation 12 can also be applied to the case of an ammonia molecule.  In a rarified atmosphere the molecule will most likely be found
in its symmetric coherent form  $\frac{1}{\sqrt{2}}(A + B)$, where $A$ has the nitrogen atom on one side of the hydrogen plane, and
$B$ has the nitrogen atom placed symmetrically on the other side.  This is the lowest energy state available to the molecule. 
Collisions with other molecules in the environment will tend to destroy this coherence, causing the ammonia molecule to become
decoherent to some extent.  This decoherence can be reversed by decreasing the pressure.  Since an ammonia molecule wants to fall into
its lowest energy level, it will tend to return to the symmetric (coherent) state when outside pressure is reduced.  In general, 
equilibrium can be found between a given environment and some degree of decoherence.  

	The ammonia molecule cannot initially assume the coherent form $\frac{1}{\sqrt{2}}(A + B)$ if the environmental collisions are too
frequent -- i.e., if the pressure exceeds about 0.5 atm.\ at room temperature \cite{JZ}.   At low pressures the molecule is a stable
coherent system, and at high pressures it is a stable decoherent system.  It seems to change from a microscopic object to a
macroscopic object as a function of its environment.  This furter supports the idea that the micro/macroscopic distinction is not
fundamental.  

\vspace{.4cm}

Equation 12 is not general enough to accommodate the main example of this paper, which is a detector that may or may not capture a
particle.  To deal with this case we replace the  coefficients $\frac{1}{\sqrt{2}}$ in eq. 12 with  time dependent
coefficients $m(t)$ and $n(t)$, where $m(t_0) = 1, n(t_0) = 0$, and where $m(t)$ decreases in time keeping
$m(t)^2 + n(t)^2 = 1$.  These coefficients describe the progress of the primary interaction, giving  
\begin{eqnarray}
\Phi(\tau > t \ge t_0) &=& [m(t)A + n(t)\underline{B}]E \rightarrow m(t)AE_A(t) + n(t)\underline{B}E_B(t) \nonumber\\
&&   \hspace{.95cm}  t = t_0  \hspace{3.3cm} t = \tau \nonumber
\end{eqnarray} 
where the transition from $A$ to $\underline{B}$ is now discontinuous \emph{and} irreversible.  This can be written
\begin{eqnarray}
\Phi(t = t_0) &=& AE \hspace{.5cm}\mbox{(intial  component)}\hspace{1.8cm} m(t_0) = 1 \\
\Phi(\tau \ge t \ge t_0) &=&  AE \rightarrow [m(t)AE^m(t) + n(t)\underline{B}E^n(t)]\hspace{.4cm} 1 \ge m(t) \ge 0  \nonumber
\end{eqnarray} 
State $\underline{B}$ in this equation is a `ready' state as
required by nRule (2); and because primary current flows from $A$ to $\underline{B}$ inside the  square bracket,
$\underline{B}$ is a candidate for state reduction according to nRule (3).  

	Equation 15 applied to our example in eq.\ 3 with $m(t)A = \psi(t)d_0$ and $n(t)\underline{B} = \underline{d}_w(t)$ says that 
initially coherent states $\psi(t)d_0$ and $\underline{d}_w$ rapidly become decoherent.  Because of the macroscopic nature of the
detector, decoherence at time $\tau$ may be assumed to be complete, and the decay time from $t_0$ to $\tau$ extremely short lived. 
The  time for any newly created pair of macroscopic objects to approach full decoherence is so brief that it is not measurable in
practice.  Still, we see that decoherence does not happen immediately when a new macroscopic state is stochastically created.
Decoherence is not  instantaneous.

\section*{Grounding the Schr\"{o}dinger Solutions}
	Traditional quantum mechanics is not completely grounded in observation inasmuch as it does not include an observer.  The
epistemological approach of Copenhagen does not give the observer a role that is sufficient for him to realize the full empirical
potential of the theory; and as a result, this model encourages bizarre speculations such as the many-world interpretation of
Everett or the cat paradox of Schr\"{o}dinger.  However, when rules are written that allow a conscious observer to be given an
ontologically complete role in the system, these empirical distortions disappear.  It is only because of the incompleteness of the
epistemological model by itself that these fanciful excursions seem plausible\footnote{Physical theory should be made to accommodate
the phenomena, not the other way around. Everett proceeds the other way around when he creates imaginary phenomenon to accommodate
standard quantum mechanics.  If the nRules (or oRules) were adopted in place of the Born rule, such flights of fantasy would not
be possible.}.

\section*{Limitation of the Born Rule}
Using the Born rule in standard theory, the observer can only record an observation at a given instant of time, and he must do so
consistently over an ensemble of observations.  Therefore, he cannot himself become part of the system for any finite period of
time.  When discussing the Zeno effect it is said that continuous observation can be simulated by rapidly increasing the number of
instantaneous observations; but of course, that is not really  continuous.    

	 On the other hand, the observer in an ontological model can \emph{only} be continuously involved with the observed system.  Once
he is on board and fully conscious of a system, the observer can certainly try to remove himself ``immediately".  However, that
effort is not likely to result in a truly instantaneous conscious observation.    So the epistemological observer claims to make
instantaneous observations but cannot make continuous ones; and the ontological observer makes continuous observations but cannot
(practically) make instantaneous ones.  Evidently the Born rule would require the ontological observer to do something that cannot be
realistically done.  Epistemologically we can ignore this difficulty, but a consistent ontology should not match a continuous
physical process with continuous observation by using a discontinuous rule of correspondence.  Therefore, an ontological model
should not make fundamental use of the Born interpretation that places unrealistic demands on an observer.

\section*{Status of the Rules}
No attempt has been made to relate conscious brain states to particular neurological configurations.  The nRules are an empirically
discovered set of macro relationships that exist on another level than microphysiology, and there is no need to connect these two
domains.  These rules preside over physiological detail in the same way that thermodynamics presides over molecular detail.  It is
desirable to eventually connect these domains as thermodynamics is now connected to molecular motion; and hopefully, this is what a
covering theory will do.  But for the present we are left to investigate the rules by themselves without the benefit of a wider
theoretical understanding of state reduction or of conscious systems.  There are two rule-sets of this kind giving us two different
quantum friendly ontologies - the nRules of this paper and the oRules of refs. 1-3.  
  
The question is, which of these two rule-sets is correct (or most correct)?  Without the availability of a wider theoretical
structure or a discriminating observation, there is no way to tell.  Reduction theories that are currently being considered may
accommodate a conscious observer, but none are fully accepted.  So the search goes on for an extension of quantum mechanics that is
sufficiently comprehensive to cover the collapse associated with an individual measurement.  I expect that any such theory will
support one of the ontological rule-sets, so these rules might server as a guide for the construction of a wider theory.

\pagebreak


\begin{thebibliography}{99}

 

\bibitem{RM1}R. A. Mould, ``Quantum Brains: The oRules"; \emph{AIP Conf. Proc.}, \textbf{750}, \mbox{p. 261} (FPP3, V\"{a}xj\"{o}
Univ., Sweden, June 2004)  
   
\bibitem{RM2}R. A. Mould, ``Quantum Brain oRules",  physics/0406016 

\bibitem{RM3}R. A. Mould, ``Quantum Brain States", \emph{Found. Phys.} \textbf{33} (4) 591-612 (2003), 

         quant-ph/0303064

\bibitem{JRF}J. R. Friedman, V. Patel, W. Chen, S. K. Tolpygo, J. E. Lukens, ``Quantum superposition of distinct macroscopic
states", \emph{Nature} \textbf{406}, 43 (2000)

\bibitem{RM5}R. A. Mould, ``The Cat nRules", quant-ph/0410147

\bibitem{DV}D. Vion, A. Aassime, A. Cottet, P. Joyez, H. Pothier, C. Urbina, D. Esteve, H. H. Devoret, ``Manipulating the Quantum
State of an Electrical Circuit'', \emph{Science}, \textbf{296}, 886 (2002)

\bibitem{YY}Y. Yu, S. Han, X. Chu, S-I. Chu, Z. Wang, ``Coherent Temporal Oscillations of Macroscopic Quantum States in a
Josephson  Junction
'', \emph{Science}, \textbf{296}, 889 (2002)

\bibitem{JZ}E. Joos, H. D. Zeh, C. Kiefer, D. Giulini, J. Kupsch, I. -O. Stamatescu, \emph{Decoherence and the Appearance of a
Classical World in Quantum Mechanics}, p. 106, Springer-Verlag, Berlin (2003)




\end{thebibliography}
\end{document}